# Machine learning for the prediction of voter model opinions through complex network structures

Aruane Mello Pineda,
*Institute of Mathematical and Computer Sciences (ICMC), University of São Paulo (USP),*
São Carlos, São Paulo, Brazil,
aruane.pineda@usp.br

Caroline Lourenço Alves,
*Institute of Mathematical and Computer Sciences (ICMC), University of São Paulo (USP),*
São Carlos, São Paulo, Brazil,
caroline.lourenco.alves@usp.br

Colm Connaughton,
*Mathematics Institute, University of Warwick,*
Coventry, England
c.p.connaughton@warwick.ac.uk

Francisco Aparecido Rodrigues
*Institute of Mathematical and Computer Science (ICMC), University of São Paulo (USP),*
São Carlos, São Paulo, Brazil,
francisco@icmc.usp.br

*Abstract*—The inference of outcomes in dynamic processes from structural features of systems is a crucial endeavor in network science. Recent research [1] has suggested a machine learning-based approach for the interpretation of dynamic patterns emerging in complex network systems. The hypothesis is applied in this study towards showing opinions can be classified in the voter model with the use of topological features from complex networks. An analysis was first performed with 2 (0 or 1) opinions from a voter model and extended for 3 and then 4 opinions from the same model. A breakdown of the key network topological features for the estimation is provided and network metrics are ranked in the order of importance with high accuracy. Our generalized approach is applicable to dynamical processes running on complex networks. This study is a step towards the application of machine learning methods to studies of dynamical patterns from complex network systems.

## INTRODUCTION

The recent popularization of online social networks has increased the relevance of the development of models and mathematical methods for the prediction of collective behavior and control of the propagation of information [2]. Models of social dynamics have been used in several fields, such as epidemics and information propagation [3–5].

The Theory of Complex Networks was introduced in the late 1990s as a means of studying the organization and dynamics of social, biological, technological, and information networks [6]. Consequently, some models, such as epidemic ones, have been adapted towards considering the structure of social networks [7] and epidemics have been modeled in networks [8], since the propagation process is strongly influenced by the structure of the network under study [2]. Moreover, all network properties are known to exert an equal impact on system dynamics, and some of them are more important than others. The mapping of the properties of a network that are more relevant for system dynamics is necessary for tuning models, obtaining more precise predictions, better understanding the consequences of topological changes in the network, and selecting interventions that will result in the best outcomes for the system.

Several models of social dynamics in complex networks such as percolation [9], Ising [10, 11], and Axelrod [12] have been studied in the scientific literature. Models related to the formation of opinions and associated social phenomena are of particular interest to the study of interactions defined through complex networks [13, 14]. Binary stochastic models have been useful for the comprehension of a variety of collective dynamics in different fields of knowledge and applied to classical problems in statistical physics, biology, and even epidemiology, including equilibrium and non-equilibrium phase transitions and spread of diseases in a population [15]. The voter model [16] consists of a simplified description of the behavior of a social network, in which network agents try to decide on a certain issue (e.g., who to vote for in an election [15, 17, 18]); it provides a simple and extensively solved example of cooperative behavior [19].

An accurate mapping of interaction patterns between the components of a system is fundamental for the description of its functionality. Such endeavors can be particularly troublesome, since the rapport between a system's structure and function is frequently not simple or linear and requires the prediction of outcomes in a dynamic system whose structure is often sparsely known. The knowledge of the consequences of structural changes in network dynamics is also fundamental for the understanding of a system's functionality. Challenges include inspecting nonlinearities in the responses of each node and finding spatial and temporal correlations derived from interconnected patterns, as well as the associated feedbacks [1].

This paper addresses an analysis of the effects of a network's structure on the evolution of social dynamics, such as classification of the voter model, towards proving whether complex networks are useful tools for the study of dynamic processes rather than exploring the voter model and its initial configuration, transient final state, and machine learning applications.

Furthermore, if the structure of the network influences the voter model (i.e. the dynamics), dynamics can be predicted by analyses of such a structure [1].

The paper also presents a general methodology for the prediction of variables associated with dynamical processes in complex network models, namely opinions available on the voter model, and investigates whether classifications can be performed with the use of the network structure from voter models with more than two opinions.

## METHODS

A dynamic variable $Y_i$ is associated with each node $i$ for the specification of the learning model. The variable is time dependent, but can reach a constant value for stationary processes, $Y_i (t \rightarrow \infty) = Y_i$. In the voter model, $Y_i$ can be an indicator variable such that $Y_i = 1$ if node $i$ has one opinion, and $0$ if it has the other opinion. Its value is related to a feature vector extracted from the structure of network $X_i = \{X_{i1}, X_{i2}, \ldots, X_{id}\}$ and that comprises measures describing structural properties, chosen specifically to classify $Y_i$. Although feature selection and model comparison algorithms can choose the elements in $X_i$, traditional network metrics were adopted in this study and the following measures were included: a) degree (K), b) clustering coefficient (C), c) closeness centrality (CC), d) betweenness centrality (B), e) Page Rank (PR), f) k-core (KC), and g) Eigenvector Centrality (EC). The goal of such models is the classification of the opinions available in them. The classification problem is related to the voter model, in which each $i$ node can be represented by an opinion of the agents (first, a 2-opinion model was used and then 3- and 4-opinion ones). The evolution dynamics in the model occurs as follows:

- an agent $i$ is randomly selected at each interaction;
- look at the opinions of neighbors ($V_i$), where $V_i$ is the set of neighboring agents of $i$;
- $i$ changes its opinion according to the opinion of the $V_i$;
- 100 steps are calculated from 10 initial configurations of the voter model and the mode classification is taken.

The voter model was applied to the four following social network datasets freely available in [20] and [21]:

- **Ego-Facebook network (an online social network with 4.039 nodes):** a dataset composed of 'circles' (or 'friends lists') from Facebook, a social network. Facebook data were collected through surveys with participants using the Facebook app. The dataset includes circles, ego networks, and node features (profiles).

- **CiaoDVD trust network (an online social network with 4.658 nodes):** the user–user trust network of site http://dvd.ciao.co.uk/ from 2013.

- **Advogato network (an online social network with 6.539 nodes):** the trust network of Advogato, an online platform for developers of free software launched in 1999. Nodes represent platform users and directed edges denote relationships of trust. A trust link is called a "certification" and divided into three different levels corresponding to three different edge weights, namely apprentice (0.6), journeyer (0.8), and master (1.0). A user with no trust certificates is called observer.

- **LastFM Asia Social Network (an online social network with 7.624 nodes):** a network for LastFM users from Asian countries, collected from the public API in March 2020. Nodes represent LastFM users from Asian countries and edges are mutual follower relationships between them. The extraction of vertex features is based on artists liked by the users.

## RESULTS

This section provides the results for the dynamical processes from simple machine learning algorithms. A set of nodes was defined in the training set for the classification and the model was adjusted according to the hold-out method for the selection of both test (25%) and training (75%) sets with the use of stratified k-folds cross-validation with k= 10 folds. Support Vector Machine (SVM) and Random forest (RF) machine learning methods were adopted for the classification. Subsequently, the opinions of the voter model were extended towards answering the following question: If the number of opinions of the model is increased to 3 and 4 opinions, is such a classification still possible?
The RF classifier applied to the 4 datasets provided an acceptable area under the ROC curve (AUC), with all values greater than 0.70. Table I shows the AUC results for 2, 3, and 4 opinions - note the values decrease as the opinions of the model increase. Overall, AUCs of 0.93 (RF) and 0.84 (SVM) were obtained for two-opinion models from the Facebook datasets.
Figures 1, 2, 3, and 4 display the importance of the measures of complex networks for our datasets - SHapley Additive exPlanations (SHAP values) show the contribution of each resource to the classification. The most important measure was EC, followed by CC.

Appendix A provides the results for two opinions of the voter model. Tables II and III show the results of RF and SVM classifiers, respectively. AUC, Accuracy, F1 score, Recall, and Precision were calculated and all values are close to 1.0, indicating high-quality classification models. The application of RF and SVM to the four databases studied enabled classification – an exception was Asia dataset, whose results were satisfactory only for RF and two-opinion models. Appendix B provides the same results reported in Appendix A, but for three opinions. Tables IV and V show the results for RF and SVM classifiers, respectively. Finally, Appendix C provides the results for RF (Table VI) and SVM (Table VII) classifiers for four opinions. Overall, regardless of the number of opinions, the AUC values are acceptable; however, when the number of opinions of the model is increased, some measures no longer produce good results. (See VI for the accuracy for LastFM Asia dataset).

TABLE I. AUC for two and more opinions of voter model and RF classifier.

| Dataset | Nodes | 2 opinions | 3 opinions | 4 opinions |
|---|---|---|---|---|
| Facebook | 4039 | 0.93 | 0.84 | 0.86 |
| DVD | 4658 | 0.94 | 0.91 | 0.85 |
| Advogato | 6539 | 0.90 | 0.80 | 0.73 |
| Asia | 7624 | 0.72 | 0.64 | 0.63 |

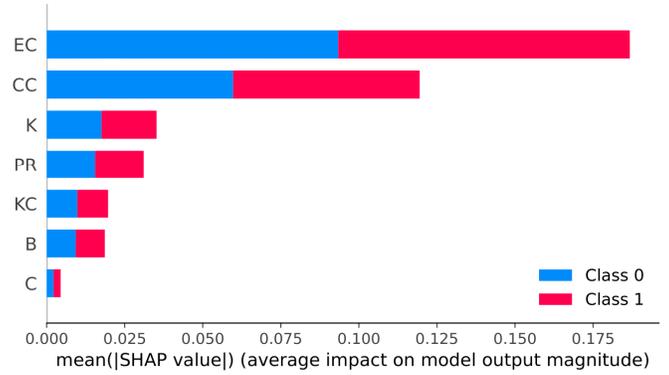

**FIG. 3.** Most important complex network measures for Advogato dataset with SHAP values.

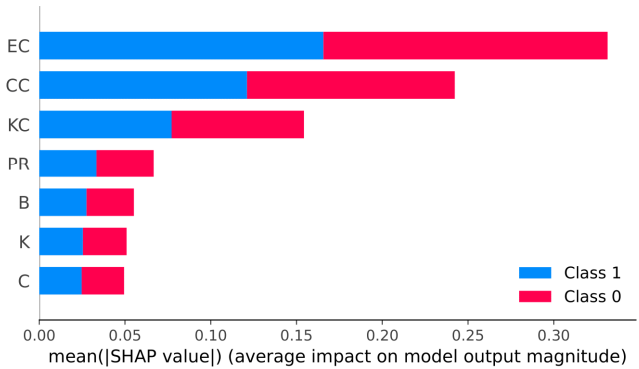

**FIG. 1.** Most important complex network measures for Facebook dataset with SHAP values.

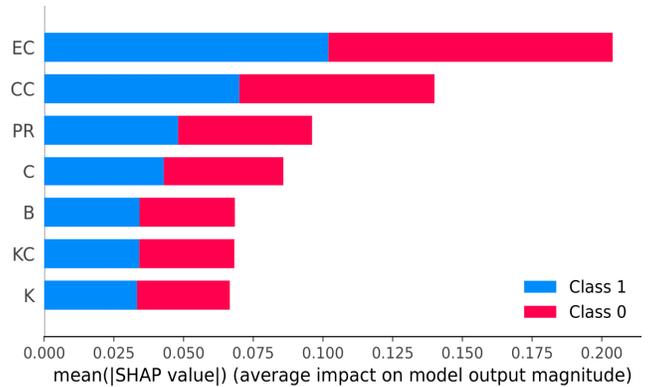

**FIG. 4.** Most important complex network measures for Asia dataset with SHAP values.

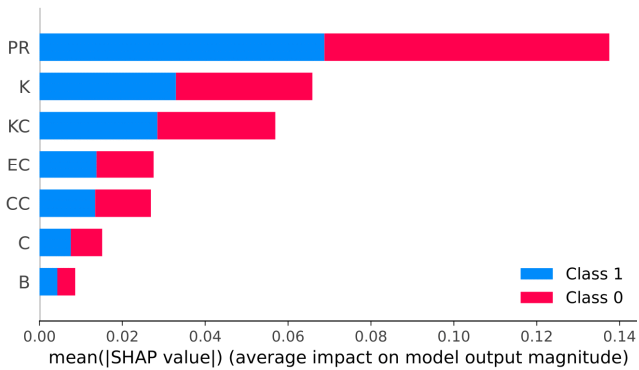

**FIG. 2.** Most important complex network measures for DVD dataset with SHAP values.

## CONCLUSIONS

This study demonstrated the feasibility of using features extracted from a subset of nodes of a complex network for estimating dynamical outcomes and introduced a methodology for the classification of opinions in the voter model. Social dynamics were simulated in complex networks according to a simple model and machine learning tools, showing the methodology is applicable to other models of social dynamics. The method is generalizable so that it can be applied to other models for predictions of a random dynamic variable from a set of nodes and their dynamics. The methodology broadens the possibilities for research on structures and dynamics of complex systems with machine learning methods.

## Appendix A: Results for two opinions

TABLE II. Results for two opinions for RF classifier.

| Dataset | nodes | AUC | Accuracy | F1 score | Recall | Precision |
|---|---|---|---|---|---|---|
| Facebook | 4039 | 0.93 | 0.93 | 0.93 | 0.86 | 0.93 |
| DVD | 4658 | 0.94 | 0.98 | 0.94 | 0.74 | 0.85 |
| Advogato | 6539 | 0.90 | 0.89 | 0.90 | 0.93 | 0.75 |
| Asia | 7624 | 0.72 | 0.73 | 0.59 | 0.72 | 0.72 |

TABLE III. Results for two opinions for SVM classifier.

| Dataset | nodes | AUC | Accuracy | F1 score | Recall | Precision |
|---|---|---|---|---|---|---|
| Facebook | 4039 | 0.86 | 0.86 | 0.86 | 0.86 | 0.86 |
| DVD | 4658 | 0.74 | 0.97 | 0.74 | 0.74 | 0.75 |
| Advogato | 6539 | 0.93 | 0.88 | 0.80 | 0.93 | 0.75 |
| Asia | 7624 | 0.59 | 0.64 | 0.59 | 0.59 | 0.61 |

## Appendix B: Results for three opinions

TABLE IV. Results for three opinions for RF classifier.

| Dataset | nodes | AUC | Accuracy | F1 score | Recall | Precision |
|---|---|---|---|---|---|---|
| Facebook | 4039 | 0.84 | 0.82 | 0.79 | 0.79 | 0.80 |
| DVD | 4658 | 0.91 | 0.85 | 0.74 | 0.74 | 0.74 |
| Advogato | 6539 | 0.80 | 0.85 | 0.48 | 0.61 | 0.44 |
| Asia | 7624 | 0.64 | 0.55 | 0.52 | 0.52 | 0.52 |

TABLE V. Results for three opinions for SVM classifier.

| Dataset | nodes | AUC | Accuracy | F1 score | Recall | Precision |
|---|---|---|---|---|---|---|
| Facebook | 4039 | 0.84 | 0.72 | 0.64 | 0.62 | 0.75 |
| DVD | 4658 | 0.88 | 0.72 | 0.44 | 0.45 | 0.54 |
| Advogato | 6539 | 0.80 | 0.85 | 0.49 | 0.63 | 0.45 |
| Asia | 7624 | 0.67 | 0.54 | 0.41 | 0.44 | 0.69 |

## Appendix C: Results for four opinions

TABLE VI. Results for four opinions for RF classifier.

| Dataset | nodes | AUC | Accuracy | F1 score | Recall | Precision |
|---|---|---|---|---|---|---|
| Facebook | 4039 | 0.86 | 0.82 | 0.78 | 0.79 | 0.78 |
| DVD | 4658 | 0.85 | 0.97 | 0.66 | 0.66 | 0.70 |
| Advogato | 6539 | 0.73 | 0.82 | 0.23 | 0.25 | 0.45 |
| Asia | 7624 | 0.63 | 0.53 | 0.45 | 0.45 | 0.45 |

TABLE VII. Results for four opinions for SVM classifier.

| Dataset | nodes | AUC | Accuracy | F1 score | Recall | Precision |
|---|---|---|---|---|---|---|
| Facebook | 4039 | 0.84 | 0.70 | 0.57 | 0.55 | 0.78 |
| DVD | 4658 | 0.66 | 0.96 | 0.33 | 0.33 | 0.33 |
| Advogato | 6539 | 0.73 | 0.83 | 0.34 | 0.48 | 0.31 |
| Asia | 7624 | 0.64 | 0.50 | 0.22 | 0.28 | 0.48 |